\documentclass[twocolumn,amsmath,amssymb,floatfix,superscriptaddress,prl,hyperref]{revtex4}
\usepackage[usenames,dvipsnames]{xcolor}
\usepackage{graphicx}
\usepackage{dcolumn}

\newcommand{\ie}{{\it i.e. }}

\newcommand{\unm}{~\mu\mathrm{m}}

\renewcommand{\vec}[1]{\mathbf{#1}}

\begin{document}

\title{Turbulent fracture surfaces: A footprint of damage percolation?}

\author{St\'ephane Vern\`ede}
\affiliation{EO Technology, 9 Hehai East Road, 213000 Changzhou, China}
\affiliation{Baseline Technical Services, 125 Jiangsu North Road , 200042 Shanghai, China}
\author{Laurent Ponson}
\affiliation{Institut Jean le Rond d'Alembert (UMR 7190), CNRS - Universit{\'e} Pierre et Marie Curie, Paris, France}
\author{Jean-Philippe Bouchaud}
\affiliation{Capital Fund Management, 23 rue de l'Universit\'e, 75007 Paris, France}

\begin{abstract}
We show that a length scale $\xi$ can be extracted from the spatial correlations of the ``steep cliffs'' that appear on fracture surface. Above $\xi$, the slope amplitudes are uncorrelated and the fracture surface is mono-affine. Below $\xi$, long-range spatial correlation lead to a multi-fractal behavior of the surface, reminiscent of turbulent flows. Our results support a unifying conjecture for the geometry 
of fracture surfaces: for scales $> \xi$ the surface is the trace left by an elastic line propagating in a random medium, while for scales $< \xi$ the highly correlated patterns on the surface result from the merging of interacting damage cavities. 
\end{abstract}

\maketitle 

After thirty years of research, it is now well established that fracture surfaces exhibit robust universal fractal statistical properties,
first reported in~\cite{Mandelbrot} and recently reviewed in~\cite{Bonamy6}.
Yet, identifying the physical mechanisms that lead to such fractal structures is still an open problem~\cite{Alava2}. The most commonly used approach to characterize the roughness of fractal cracks is to study the scaling of the off-plane height variation $\delta h$ of the fracture surface with the observation scale $\delta r$. The variance of this distribution shows a scaling law  $\langle \delta h^2 \rangle \sim \delta r^{2\zeta}$ where $\zeta$ is the so-called roughness exponent. For purely brittle failure, the roughness exponent is reported to be $\zeta \approx 0.45$~\cite{Boffa,Ponson6} whereas for materials that undergo damage during failure, $\zeta \approx 0.75$~\cite{Bouchaud9,Maloy}. It has been conjectured that these exponents are the signature of the fracture mechanism above and below the size of the process zone~\cite{Bonamy2}. However, standard methods for extracting roughness exponents are not able to elicit the differences between the fracture mechanisms in the two regimes.

Here we propose a different approach for characterizing crack roughness statistics by focusing on the {\it local slopes} of the fracture surfaces and their spatial correlations. This allows us to identify unambiguously two scaling regimes: above some length scale $\xi$, the slope amplitudes are uncorrelated and the fracture surface displays a mono-affine Gaussian behavior with a roughness exponent of  $\zeta \approx 0.45$. Below $\xi$, long-range spatial correlations do appear and lead to a multi-fractal behavior of the surface. Our findings show that the presence of two distinct regimes of roughness first reported in Refs.~\cite{Morel8,Santucci} is a generic feature of fracture surfaces and is reminiscient of the brittle mode of failure that takes place at large scale and of the damage mechanisms present in the tip vicinity. In addition, it reveals the subtle organization of crack roughness at small length scales $\delta x < \xi$, reminiscent of the phenomenology of turbulent flows~\cite{Frisch}. In particular, we relate quantitatively the multi-fractal spectrum measured at these length scales with the spatial correlations of the local slopes, and show that the largest slopes organize into a network of lines or ``steep cliffs'' that exhibit universal statistics. This new approach to the characterisation of fracture surfaces brings insights on the microscopic mechanisms at play during material failure, in particular on the mechanism of damage percolation taking place at the tip of cracks. It also paves the way to a post-mortem measurement of the size of the crack tip damaged zone, as a promising tool to infer material toughness from the statistical analysis of fracture surfaces.

\begin{figure}
\includegraphics[width=1\columnwidth]{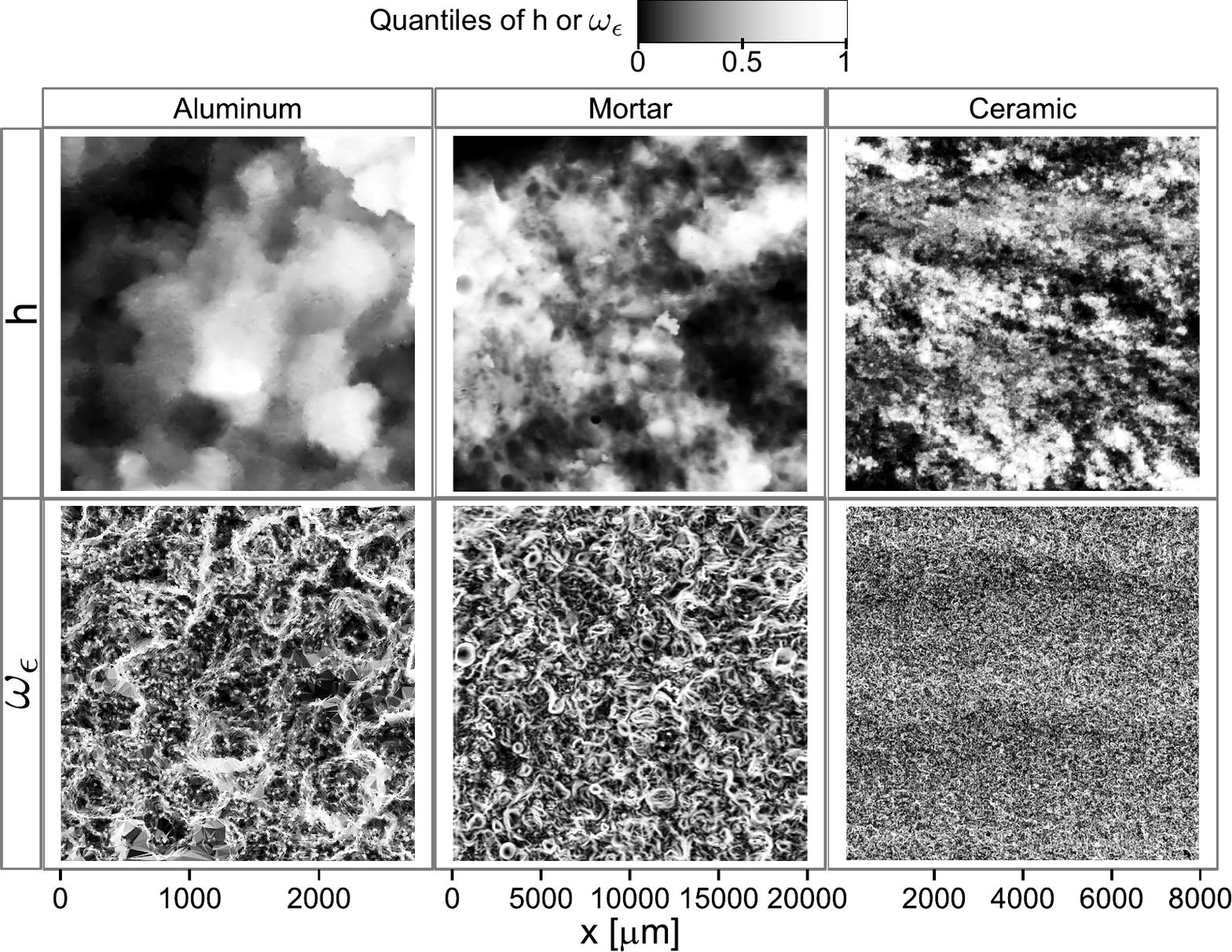}
\caption{Maps of $h$ and $\omega_\epsilon$ for the three materials studied. Top: $h$, height of the measured fracture surface. Bottom: $\omega_\epsilon$ transformation providing the field of local slopes computed at a scale $\epsilon$. In both cases, the quantiles of the distribution are represented by a grey hue, the largest values being represented by the lightest hue. $\omega_\epsilon$ is computed at the scales $\epsilon= 3$, $50$ and $8\unm$ for the aluminum, mortar and ceramic fracture surfaces, respectively. }
\label{Fig_Maps}
\end{figure}
For this work, we have selected three sample materials that show a wide range of fracture behavior, namely an aluminum alloy, a mortar, and a sintered glass beads ceramic. The aluminum alloy specimens are aluminum 4 wt$\%$ copper broken under uni-axial mode I tension at $620^{\circ}\mathrm{C}$, in a semi-solid state~\cite{Vernede2}. The fracture surfaces are observed with a scanning electron microscope at two tilt angles and the elevation map is produced from a cross-correlation surface reconstruction technique. The mortar fracture surface is obtained by applying four points bending under controlled displacement conditions to a notched beam~\cite{Morel8}. The topography of the fracture surfaces is recorded using an optical profilometer. The sintered glass beads ceramic fracture surfaces are obtained with a tapered double cantilever beam broken at constant opening rate~\cite{Ponson6}. The roughness of the fractured specimen is measured using a mechanical stylus profilometer. Those fracture surfaces are described by their height field $h(\vec{x})$, function of a two-dimensional in-plane vector $\vec x$, that are represented in a grayscale on the top panels of Fig.~\ref{Fig_Maps} for each material.

\begin{figure}
\includegraphics[width=1\columnwidth]{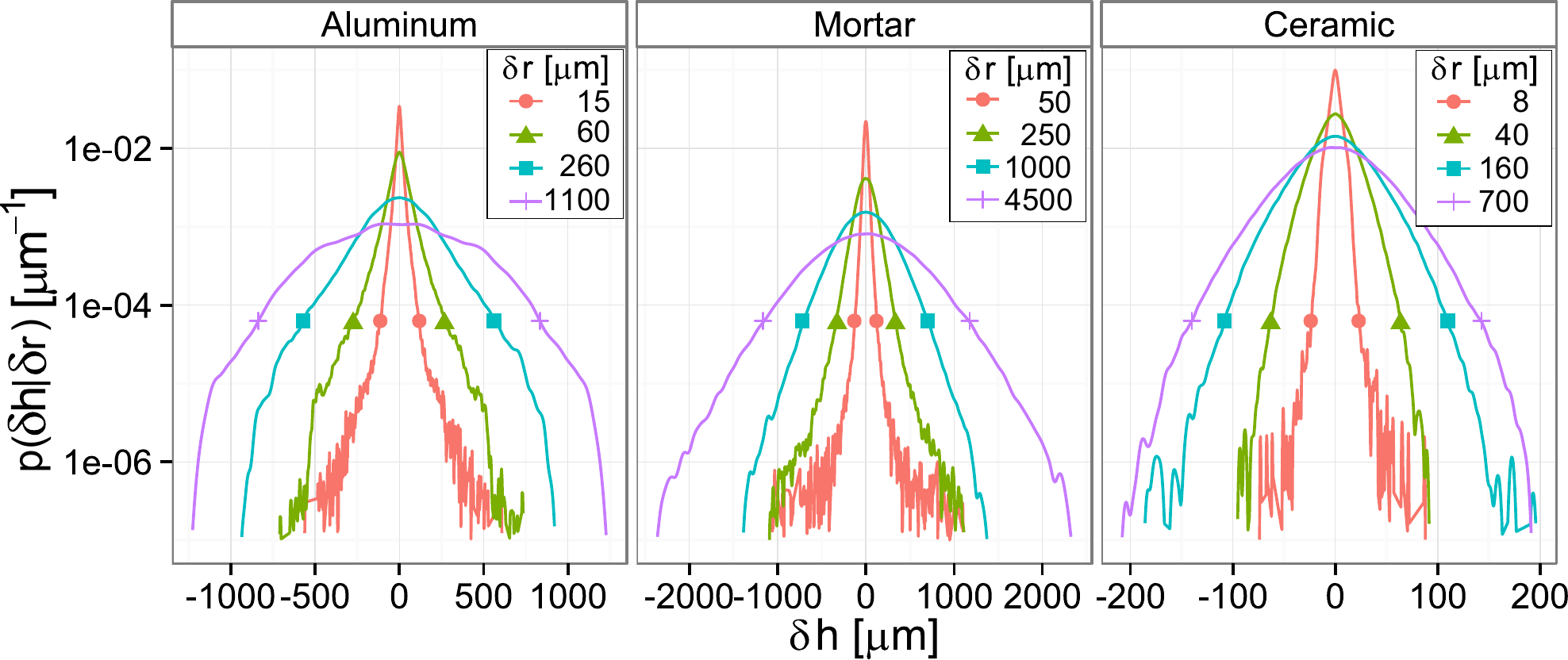}
\caption{Distribution of height fluctuations $p(\delta h|\delta r)$ at various scales $\delta r$ for the three samples considered.}
\label{Fig_Distribs}
\end{figure}

A first natural step in the characterization of the roughness statistics is to compute the distribution of height fluctuations at different scales. For a given increment $\delta \vec{x}$ of the coordinates in the average fracture plane, we note $p(\delta h|\delta \vec{x})$ the probability distribution of an height increment $\delta h = h(\vec{x}) - h(\vec{x}+ \delta \vec{x})$ where the sampling of the distribution is done on all admissible coordinates $\vec{x}$. We also note $p(\delta h|\delta r)$ the distribution of $\delta h$ where the sampling is done on all admissible $\vec{x}$ and $\delta \vec{x}$ such as $|\delta \vec{x}|=\delta r$. The distribution $p(\delta h|\delta r)$ at different $\delta r$ is shows in a semi-logarithmic scale in Fig.~\ref{Fig_Distribs} for aluminum, mortar and ceramics fracture surface. In this semi-logarithmic representation, the parabolic shape of the distribution obtained for large values of $\delta r$ reveals Gaussian statistics. This contrasts with the distributions observed for smaller values of $\delta r$ that display fat tails. This drastic change in the shape of the distribution indicates thata single exponent is unsufficient to describe the variations of the roughness properties with the scale of observation. Fat tail statistics also suggest the presence of over-represented large height variations over small in-plane distances that we would like to analyze further.


To investigate the spatial distribution of these steep ``cliffs", we introduce the quantity $\omega_{\epsilon}(\vec x)$ that measures the intensity of the local height variations of the fracture surface on a scale $\epsilon$:
\begin{equation}
\omega_{\epsilon}(\vec{x}) = \frac12 \log \left(\langle \delta h(\vec{x},\delta \vec{x})^2\rangle_{|\delta \vec{x}|=\epsilon}\right) - \Omega_\epsilon.
\label{Eq_w}
\end{equation} 
$\delta h(\vec{x},\delta \vec{x})= h(\vec{x}+\delta \vec{x}) -  h(\vec{x})$ is the local slope of the surface in the direction $\delta \vec{x}$, and $\Omega_\epsilon$ is chosen such that the average of $\omega_{\epsilon}(\vec{x})$ over all $\vec x$ is zero. Note that the average of the slopes is done over a circle of radius $\epsilon$. This new field $\omega_{\epsilon}(\vec{x})$ has several interesting properties, like isotropy and robustness to measurement artifacts. The maps of $\omega_\epsilon$ calculated from the off-plane height maps $h$ shown in Fig.~\ref{Fig_Maps} are represented on the lower panels in the same figure. Strikingly, the largest values of $\omega_\epsilon$ (lighter grey), corresponding to the steep cliffs that populates the tails of the distribution $p(\delta h|\delta r)$, are spatially correlated and form a network of rough lines for the aluminum and the mortar fracture surfaces. For the ceramic fracture surface, smaller patterns are visible. 

\begin{figure}
\includegraphics[width=1\columnwidth]{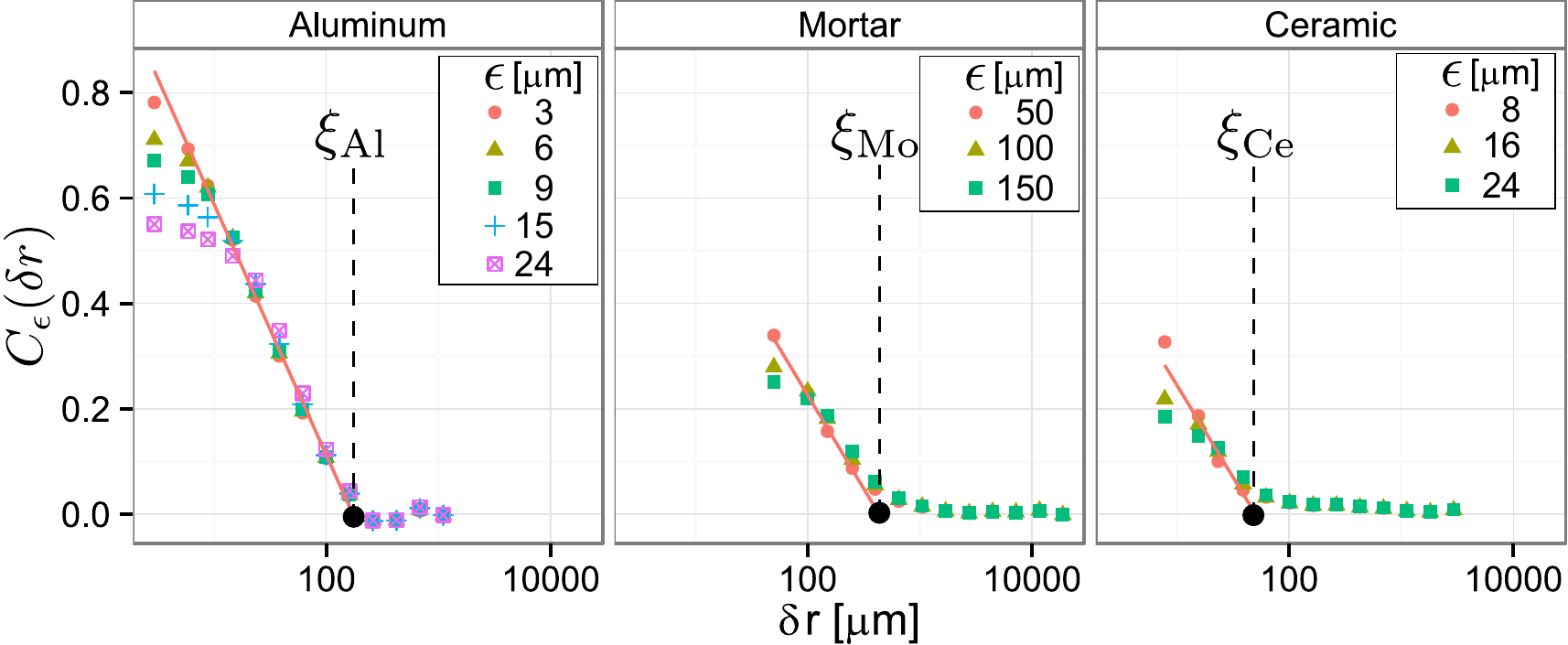}
\caption{Spatial correlations of $\omega_{\epsilon}$ for the three materials considered. The correlations are represented for $\omega_\epsilon$ computed at different scales $\epsilon$. The cut-off length $\xi$ is represented for each case.}
\label{Fig_Correlations}
\end{figure}

The visually correlated patterns in Fig.~\ref{Fig_Maps} can be quantified by computing the spatial correlations of $\omega_{\epsilon}$, which we further average over all directions~\cite{Footnote1}, \ie $C_\epsilon(\delta r) = \langle \omega_{\epsilon}(\mathbf{x})\omega_{\epsilon}(\mathbf{x}+\delta \mathbf{x}) \rangle_{\mathbf{x},|\delta \mathbf{x}|=\delta r}$. This quantity is shown in Fig.~\ref{Fig_Correlations} as a function of the distance $\delta r$ for different observation scales $\epsilon$; $C_\epsilon(\delta r)$ is independent of $\epsilon$ whenever $\epsilon \ll \delta r$. For the three materials considered, we clearly observe two regimes. At small $\delta r$, $\omega_\epsilon$ shows strong spatial correlations, which decay {\it logarithmically} with distance, \ie $C_\epsilon(\delta r) \sim - \lambda \log(\delta r/\xi)$, and extrapolates to zero for $\delta r=\xi$. For larger distances, these correlations are zero within statistical noise. Both $\lambda$ and $\xi$ are found to be, to a good approximation, independent of $\epsilon$ provided $\epsilon \ll \xi$. Note that $\lambda$ is a dimensionless number and its value is empirically found to be quite similar for the three materials at hand: $0.21 \pm 0.02$ (aluminum), $0.15 \pm 0.03$ (mortar) and $0.15 \pm 0.06$ (ceramics).  The crossover length $\xi$ is found to be $170 \pm 12$, $430 \pm 35$ and $50 \pm 9\unm$, respectively. Note that the ratio of $\xi$ to the total map size is $0.06, 0.02$ and $0.006$, respectively. These last values confirms the visual impression conveyed by Fig.~\ref{Fig_Maps} where large correlated patterns are observed for aluminum, smaller patterns for mortar and even smaller ones for ceramics. 

\begin{figure}
\includegraphics[width=1\columnwidth]{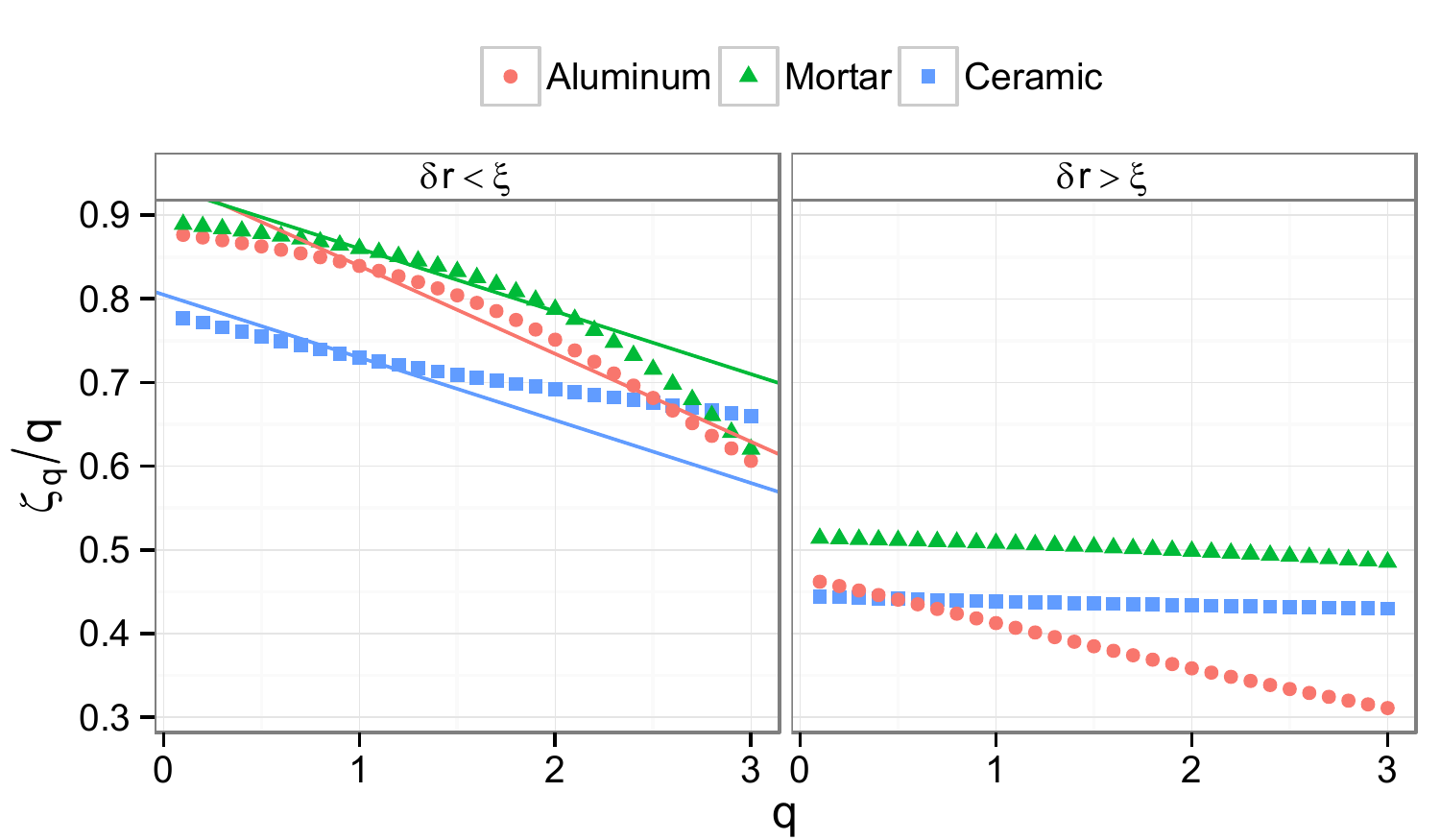}
\caption{Multifractal spectrum of the fracture surfaces height differences. The spectrum are computed both below (left) and above $\xi$ (right). The spectrum predicted for $\delta r < \xi$ by the multifractal model of Eq.~(\ref{Eq_Multiscaling}) is represented by a straight line. } 
\label{Fig_Spectrum}
\end{figure}

In order to characterize further the two regimes, we now compute the {\it multifractal spectrum} of the height fluctuations, defined through $\langle \left| \delta h(\vec{x},\delta \vec{x}) \right|^q \rangle_{\vec{x},|\delta \vec{x}|=\delta r} \sim \delta r^{\zeta_q}$ for the two range of length scales $\delta r < \xi$ and $\delta r > \xi$. Note that the standard roughness exponent $\zeta$ corresponds to $q=2$, and $\zeta\equiv\zeta_2/2$. For  $\delta r > \xi$, we observe that $\zeta_q/q$ is fairly independent of $q$ with a value around $0.45$ (Fig.~\ref{Fig_Spectrum}\,(right)). This corresponds to a mono-affine behavior, \ie a scaling that preserves the shape of the full distribution $p(\delta h|\delta r)$ of height fluctuations. This is consistent with the observation of a conserved Gaussian distribution at large scales (see Fig.~\ref{Fig_Distribs}), and in agreement with previous findings~\cite{Ponson11,Santucci,Ponson12}. The mono-affine behavior is very clear for the mortar and the ceramic fracture surface. For the aluminum fracture surface, some residual variations of $\zeta_q/q$ with $q$ is observed; this behavior may be due to the rather limited extension of the large scale regime $\delta r > \xi$.

For $\delta r < \xi$, on the other hand, we do observe a significant variation of $\zeta_q/q$ with $q$. This multi-affine behavior can in fact be traced back to the logarithmic decay of the spatial correlations $C_\epsilon(\delta r)$ of slopes discussed above. Indeed, assuming that $\omega_\epsilon$ is a Gaussian field, and that the local slope can be written as $\delta h(\vec x,\delta \vec x) = e^{\omega_\epsilon(\vec x)} s_\epsilon(\vec x)$ with $|\delta \vec{x}|=\epsilon$ and $s_\epsilon(\vec x)$ is a {\it long ranged correlated} random variable with unit variance and $\langle s_\epsilon(\vec x) s_\epsilon(\vec x + \delta \vec y) \rangle_{\vec{x},|\delta \vec{y}|=\delta r} \sim |\delta r|^{-\gamma}$, one derives,  adapting the calculations of~\cite{Muzy} (see the Supplementary Material~\cite{SI}):
\begin{equation}
\zeta_q = q \left(H-(q-1)\frac{\lambda}{2} \right) \quad \mathrm{with} \quad H \equiv \zeta_1 = (1 - \frac{\gamma}{2})
\label{Eq_Multiscaling}
\end{equation}
where $\lambda$ is the slope of the logarithmic correlation defined above. As seen in Fig.~\ref{Fig_Spectrum}\,(left), the slope of the multi-fractal spectrum is indeed well captured by this simple model. We have measured the exponent $\gamma$ independently, from the spatial correlations of the {\it signs} of the local slopes in a given direction, with good agreement with the direct estimate of $H$, in particular for aluminum where the scaling region is large. We therefore claim that fracture surfaces are, on short length scales, bi-dimensional realizations of {\it multifractal, persistent} Brownian motions. Whereas natural realizations of multifractal Brownian motions with $H \leq 1/2$ have been reported in turbulent flows ($H \approx 1/3$)~\cite{Frisch} and in financial time series ($H \approx 1/2$)~\cite{Muzy}, it is to our knowledge the first time that a multifractal signal with $H > 1/2$ is observed. The curvature of the multi-fractal spectrum seen in Fig.~\ref{Fig_Spectrum} cannot be captured by Eq.~(\ref{Eq_Multiscaling}). This can be traced back to the assumption that $\omega_\epsilon$ is a Gaussian field. Introducing non-trivial higher order correlations of $\omega_\epsilon$ that also decay logarithmically would add higher order contributions to $\zeta_q$. However, these higher order correlations are difficult to measure and we lack statistics to test the model beyond the second order correlations reported here~\cite{Footnote2}.

\begin{figure}
\includegraphics[width=1\columnwidth]{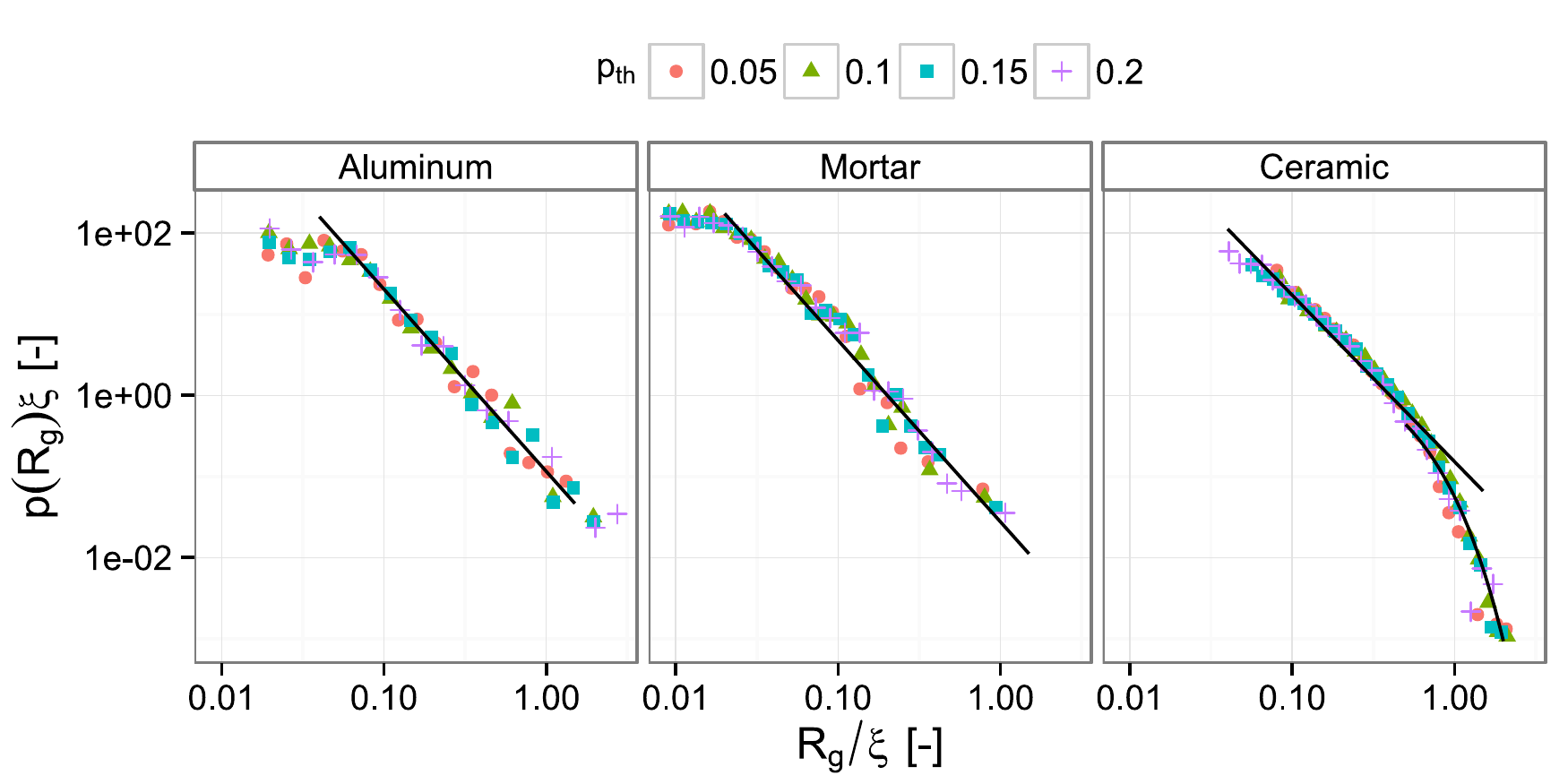}
\caption{Distribution of cluster sizes $R_\mathrm{g}$ for different thresholds $p_{\mathrm{th}}$. }
\label{Fig_Clusters}
\end{figure}

To characterize further the spatial organization of the steepest regions and its robustness towards material specificity, we study the geometrical properties of the clusters formed by the largest values of $\omega_\epsilon$, \ie the ridge of the cliffs. The field $\omega_\epsilon$ is thresholded in order to keep only a fraction $p_{\mathrm{th}}$. These extreme events organize in a network of disconnected clusters, as suggested by Fig.~\ref{Fig_Maps}\,(bottom) (see also Fig.~\ref{Fig_Maps_SI} of the Supplementary Material~\cite{SI}). The extension $\ell$ of each cluster can be defined using either its extension along the horizontal or vertical axis, or its radius of gyration $R_\mathrm{g}$. These three quantities are found to follow the same scaling $S \sim \ell^D$ with the number of pixels (or area) $S$ contained by the cluster, suggesting that these clusters have fractal geometry with dimension $D \approx 1.65 \pm 0.15$, again independently of the material considered (see Supplementary Material~\cite{SI}). Using the values of $R_\mathrm{g}$, we compute on Fig.~\ref{Fig_Clusters} the distribution of cluster sizes for different values of the threshold $p_{\mathrm{th}}$. For scales smaller than $\xi$, we observe a power law distribution $P(R_g) \sim R_g^{-\tau}$ with exponent $\tau =2.2 \pm 0.2$ for all three materials
This means that not only the roughness exponent $\zeta$ but at least {\it three} other quantities describing the statistics of fracture surfaces are universal in the small $\delta r$ regime: $\lambda$ that describes the spatial correlation of the amplitude of slopes and the multifractal spectrum, $D$ that is the fractal dimension of the ridge of the cliffs on the fracture surface and $\tau$ that characterizes the cluster sizes distribution (see Table~\ref{Tab1} for a comparison of their value from one material to another). This extended universality is important for at least two reasons: (i) it provides additional support for the conjecture that the statistics of fracture surfaces is universal, suggesting a common underlying roughening mechanism; (ii) it provides important further constraints that must be abided by any theory attempting to explain the universal value of the roughness exponent $\zeta \approx 0.75$~\cite{Bouchaud9,Maloy}.

\begin{table}
\begin{center}
\begin{tabular}{|l|c|c|c|c|c|c|c}
\hline
& $\lambda $ & $D$ & $\alpha$ & $\xi$ \\ \hline
aluminum & $0.21 \pm 0.01$ & $1.70 \pm 0.10$ & $2.2 \pm 0.2$ & $170 \pm 8\unm$ \\
Mortar & $0.15 \pm 0.01$ & $1.68 \pm 0.08$ & $2.2 \pm 0.1$ & $430 \pm 35\unm$ \\
Ceramics & $0.15 \pm 0.03$ & $1.63 \pm 0.12$ & $2.1 \pm 0.1$ & $50 \pm 8 \unm$ \\ \hline
\end{tabular}
\caption{Statistical parameters extracted from the fracture surfaces (see text for details).}
\label{Tab1}
\end{center}
\end{table}

What is the physical interpretation of these observations? We believe that the presence of spatially correlated steep cliffs on short length scales $\delta r < \xi$ is a strong indication that at these scales, fracture proceeds through nucleation and coalescence of microcracks or damage cavities, as previously conjectured~\cite{Paun,RaviChandar,Fisher,Hansen}. As elaborated in the context of planar cracks~\cite{Santucci7,Gjerden2}, the scale $\xi$ therefore provides the extent $\ell_{\mathrm{pz}}$ of the fracture process zone. It also reflects the material toughness, since $K_{\mathrm{Ic}} \simeq \sigma_{c} \sqrt{\ell_{\mathrm{pz}}}$ where $\sigma_{c}$ is the typical failure stress under tension~\cite{Barenblatt}. Our methodology based on the correlation of slopes measured on the fracture surfaces paves the way for a post-mortem characterization of material toughness from the statistical analysis of their fragments.

Although this is still an open theoretical issue, the percolation of power law distributed microcracks should provide a rationale for the value of the small scale roughness exponent $\zeta \approx 0.75$. Discontinuities along the fracture surface are the stigma of these coalescing cavities~\cite{Guerra} and this picture should be made compatible with values of the new, universal statistical indicators reported here, namely $\lambda$, $D$ and $\tau$. Only on large length scales $\delta r > \xi$ does the notion of a continuous fracture line make sense. Continuum fracture mechanics based models describing crack fronts as an elastic interface driven in a random medium~\cite{Gao,JPBouchaud,Ramanathan,Schmittbuhl4,Fisher,Bonamy2,Ponson12} predict mono-affine Gaussian fracture surfaces with $\zeta \approx 0.4$~\cite{Bonamy2,Ponson12}, indeed compatible with our findings.

 We thank D. Bonamy, E. Bouchaud, Y. Cao, A. Hansen and S. Morel for fruitful discussions. The support from the European Union through the Marie Curie Integration Grant ``ToughBridge" is gratefully acknowledged (LP).

\section{Generalization of the Bacry-Delour-Muzy analysis for long-range correlated processes}
For simplicity, we only consider here one dimensional profiles $h(x)$. We write the slope $\delta h$ on scale $\epsilon$ as $\delta h(x) = s(x) e^{\omega(x)}$, 
where $s(x)$ has long-range correlations described by the exponent $\gamma$ and $\omega(x)$ is a Gaussian field with logarithmic corrections with slope $-\lambda$. Then, by definition,
\begin{equation}
h(x) - h(0) = \sum_{i=1}^{x/\epsilon} \delta h(x_i), \qquad x_i:=i \epsilon.
\end{equation}
We want to compute the various moments of $h(x)-h(0)$. For even values of $q=2m$, and assuming for simplicity that $s(x)$ is Gaussian that allows one to use
Wick's theorem, one has:
\begin{eqnarray} \nonumber 
& &\left \langle (h(x) - h(0))^{q} \right \rangle  = C_{2m}^m \\ \nonumber
& & \sum_{i_1, \dots, i_{2m}} \langle s(x_1) s(x_2) \rangle \dots \langle s(x_{2m-1}) s(x_{2m}) \rangle
\langle e^{\sum_{\alpha = 1}^{2m} \omega(x_\alpha)} \rangle.
\end{eqnarray}
Now, since the $\omega$'s are also Gaussian with a logarithmic correlation function, one has:
\begin{equation}
\langle e^{\sum_{\alpha = 1}^{2m} \omega(x_\alpha)} \rangle = C e^{-\frac{\lambda}{2} \sum_{\alpha \neq \beta = 1}^{2m} \ln |x_\alpha - x_\beta|},
\end{equation}
where $C$ is a constant that accounts for the diagonal terms $\alpha = \beta$, that plays no role in the following. 

Let us first consider the Bacry-Delour-Muzy (BDM) case where $s(x)$ has only short range correlation, say even purely local: $\langle s(x_1) s(x_2) \rangle = \delta_{x_1,x_2}$.
This local contribution will always exist, even in the long-range case, and will eventually have to be compared to the latter contribution in order to determine the dominant one. In the BDM case, this imposes $x_{2j+1}=x_{2j+2}$, and therefore:
\begin{equation}
e^{-\frac{\lambda}{2} \sum_{\alpha \neq \beta = 1}^{2m} \ln |x_\alpha - x_\beta|} \propto e^{-2\lambda \sum_{j \neq k = 1}^{m} \ln |x_{2j} - x_{2k}|}.
\end{equation}
Hence, approximating discrete sums by integrals in the limit $\epsilon \to 0$, one has:
\begin{eqnarray} \nonumber 
& &\left \langle (h(x) - h(0))^{q} \right \rangle \propto \\ \nonumber
& &\int_0^x \dots \int_0^x {\rm d}x_2 \dots {\rm d}x_{2m} \prod_{j \neq k = 1}^{m} |x_{2j} - x_{2k}|^{-2\lambda}.
\end{eqnarray}
Setting $x_{2j} = x \, u_{2j}$, one finds by simple power-counting (which is correct provided the resulting integral on $u$'s converges, i.e. when $\zeta_q > 0$; see \cite{Muzy}):
\begin{equation}
\left \langle (h(x) - h(0))^{q} \right \rangle \propto x^m \times x^{2\lambda m (m-1)} \equiv x^{\zeta_q},
\end{equation}
or $\zeta_q = q/2 (1- \lambda (q-2))$, which is precisely the BDM result \cite{Muzy}.   

Now, if one assumes that for large distances $\langle s(x_1) s(x_2) \rangle \propto |x_1 - x_2|^{-\gamma}$, the corresponding contribution to the $q$-th moment
of $h(x)-h(0)$ reads:
\begin{eqnarray} \nonumber
& &\left \langle (h(x) - h(0))^{q} \right \rangle \propto \\ \nonumber
& & \int_0^x \dots \int_0^x {\rm d}x_1{\rm d}x_2 \dots {\rm d}x_{q} \prod_{\ell=0,m-1} 
|x_{2\ell+1} - x_{2\ell}|^{-\gamma} \\ \nonumber
& &\prod_{j \neq k = 1}^{q} |x_{j} - x_{k}|^{-\lambda/2}
\end{eqnarray}
Power-counting now leads to:
\begin{equation}
\zeta_q = q - q\frac{\gamma}{2} - q(q-1)\frac{\lambda}{2} \equiv q (H - (q-1)\frac{\lambda}{2}), \qquad H = 1 - \frac{\gamma}{2},
\end{equation}
as given in the text. However, one has to compare this last contribution to the BMD one, leading to the following inequality:
\begin{equation}
q (1- \frac{\gamma}{2} - (q-1)\frac{\lambda}{2}) \geq \frac{q}{2}(1- \lambda (q-2)) \longrightarrow \gamma < 1 -\lambda,
\end{equation}
independently of $q$. This last condition is well satisfied in practice, since $\gamma \approx 0.4-0.5$ and $\lambda \approx 0.15-0.2$.

\section{Statistical characterization of clusters formed by the largest value of $\omega_\epsilon$}
\begin{figure}
\includegraphics[width=1\columnwidth]{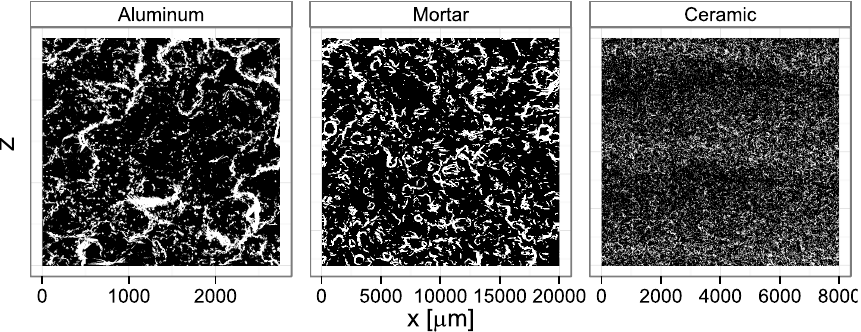}
\caption{Clusters formed by the largest values of $\omega_{\epsilon}$ for a threshold $p_{\mathrm{th}}=0.2$ for the three materials studied.}
\label{Fig_Maps_SI}
\end{figure}
The distributions of height fluctuations in the materials investigated do not follow a mono-affine Gaussian behavior at small scale because of the extended fat tails of $p(\delta h|\delta r)$. As the observation scale $\delta r$ is decreased, these tails indeed become more pronounced, resulting in the multi-affine behavior revealed by the spectrum $\zeta_q$ (see Fig.~\ref{Fig_Spectrum}). To proceed to a quantitative analysis of the spatial distribution of the steepest slopes that contribute to the distribution tails, we introduce $\omega_{\epsilon}(\vec x)$ defined by Eq.~(\ref{Eq_w}) as a transformation of the original fracture map that allows for a straightforward localization of these extreme events (see Fig.~\ref{Fig_Maps_SI}). As the focus is put on these steepest slopes, the new maps $\omega_{\epsilon}(\vec x)$ are thresholded, and the fraction $p_\mathrm{th}$ of the total number of pixels of the original map is conserved. We choose $p_\mathrm{th}$ in the range $0.05 - 0.25$. The position of the steepest slopes is represented on Extended Data Fig.~\ref{Fig_Maps_SI} for $p_{\mathrm{th}}=0.20$ on the aluminum, mortar and ceramics fracture surfaces.

\begin{figure}
\includegraphics[width=\columnwidth]{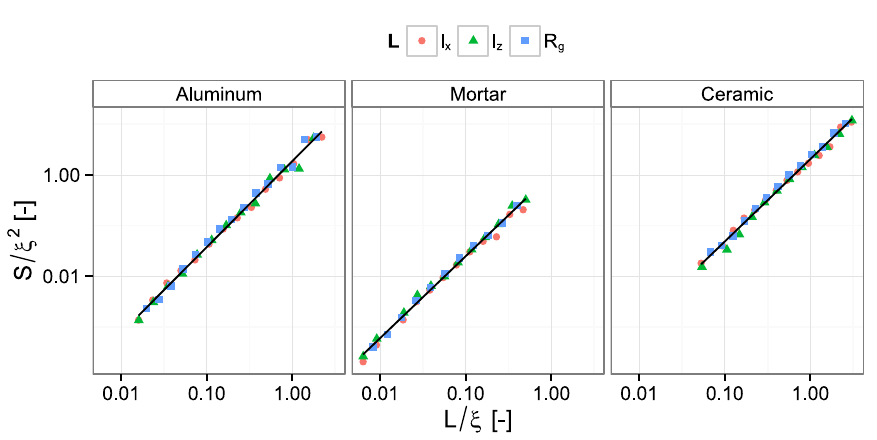}
\caption{Cluster surface as a function of its typical length scale $\ell$ for the three samples considered. The length $\ell$ is either the cluster radius of gyration $R_\mathrm{g}$, or its extension along the horizontal  $\ell_\mathrm{x}$ and vertical axis  $\ell_\mathrm{z}$. All lengths are normalized by the crossover length $\xi$.}
\label{Fig_Clusters1_SI}
\end{figure}

The statistical characterization of the geometry of these clusters is now detailed. We investigate first the geometry of the isolated clusters. Figure~\ref{Fig_Clusters1_SI} shows the variation of cluster area as a function of three typical length scales characterizing the cluster size, namely its extension $\ell_\mathrm{x}$ along the propagation direction, its extension $\ell_\mathrm{z}$ perpendicular to it and its radius of gyration $R_\mathrm{g}$. These three quantities characterize equivalently the fractal geometry of the clusters since they all scale as $\sim S^D$, where the fractal dimension $D \simeq 1.65 \pm 0.15$ irrespective of the material considered. The distributions of cluster sizes are characterized using their radius of gyration $R_\mathrm{g}$. They are shown on Fig.~\ref{Fig_Clusters} of the main article, and follow power laws with exponent $\tau \simeq 2.2 \pm 0.2$.

\begin{figure}
\includegraphics[width=1\columnwidth]{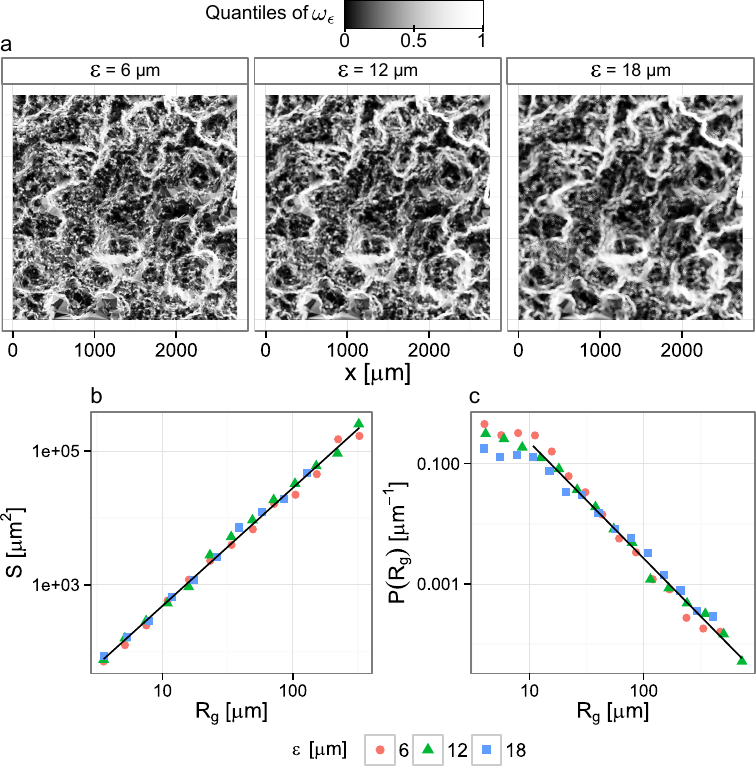}
\caption{Effect of the scale $\epsilon$ on the cluster properties for the aluminum fracture surface: (a)  $\omega_\epsilon$ field; (b) fractal geometry of the clusters; (c) cluster size distribution.}
\label{Fig_Clusters2_SI}
\end{figure}

We investigate now the robustness of these findings towards our statistical procedure. The role of $p_\mathrm{th}$ on the cluster statistics is investigated on Fig.~\ref{Fig_Clusters} of the main article, and does not show any noticeable effect. Similarly, we observe that $p_\mathrm{th}$ has a negligible effect on the cluster fractal dimension. The effect of the scale $\epsilon$ on the cluster statistical characterization is illustrated on Fig.~\ref{Fig_Clusters2_SI} in the case of the aluminum fracture surface. The clusters display robust statistical properties independent of the length $\epsilon$ as long as one considers sufficiently large cluster sizes $R_\mathrm{g} \gg \epsilon$.


\begin{thebibliography}{26}
\bibitem{Footnote1}
\bibinfo{text}{We assume, for simplicity, that fracture surfaces are statistically isotropic. This is only approximately correct, since one expects the roughness in the direction parallel to the crack propagation and perpendicular to it to be different~\cite{Ponson5}. The extension of our results to a multifractal anisotropic Family-Vicsek scaling is left for a future study.}
\bibitem{Footnote2}
\bibinfo{text}{Attempts to measure the third order correlations of the $\omega_\epsilon$ field however leads to the correct sign of the curvature of $\zeta_q/q$.}
\end{thebibliography}
\end{document}